\documentclass[12pt]{article}
\usepackage{amssymb}
\usepackage{epsfig}

\newcommand{\beqn}{\begin{eqnarray}}
\newcommand{\eeqn}{\end{eqnarray}}
\newcommand{\beq}{\begin{equation}}
\newcommand{\eeq}{\end{equation}}
\newcommand{\eq}[1]{(\ref{#1})}
\newcommand{\abstracts}[1]{{\centering{\begin{minipage}{13.0truecm}
 \normalsize\baselineskip=15pt \centerline{\footnotesize
 ABSTRACT}\vspace*{0.3cm} \parindent=20pt #1 \end{minipage}}\par}}

\newcommand{\tr}{\mbox{Tr}}

\begin{document}
~ \vspace{-1cm}
\begin{flushright}
{ ITEP-LAT/2004-03}

\vspace{0.2cm}

{ MPP-2004-3}

\end{flushright}

\begin{center}
{\baselineskip=24pt {\Large \bf Properties of P-vortex
and monopole clusters in lattice $SU(2)$ gauge theory}\\

\vspace{1cm}

{\large A.\,V.\,Kovalenko$^{\dag}$,
    M.\,I.\,Polikarpov$^{\dag}$,
    S.\,N.\,Syritsyn$^{\dag}$ and
    V.\,I.\,Zakharov$^{*}$} } \vspace{.5cm} {\baselineskip=16pt { \it

$^{\dag}$ Institute of Theoretical and  Experimental Physics,
B.~Cheremushkinskaya~25, Moscow, 117259, Russia\\
$^{*}$ Max-Planck Institut f\"ur Physik, F\"ohringer Ring 6, 80805, M\"unchen,
Germany} }
\end{center}

\vspace{5mm}

\abstracts{We study the action and geometry of P-vortices, discriminating
between the
percolating and finite clusters. We also discuss
the interrelation of the monopoles and P-vortices.
To define P-vortices we use both
the direct maximal
center projection and  indirect maximal center projection.
We find, in particular, that the action density of the P-vortices in
short clusters is substantially higher than in the percolating cluster.
The surface of the percolating cluster appears random at short distances,
with action density depending on the shape.}

\date{}

\section{Introduction}
The monopole mechanism and P-vortex mechanism are the most popular explanations
of the confinement of color \cite{reviews}. Mostly, these mechanisms are viewed
as alternatives. However, at a closer look the field configurations
representing monopoles and P-vortices turn to be interrelated
\cite{mon_pvort,deforcrand,InterplayLat}. In more detail, the monopole currents
form closed lines on 4D lattice, while P-vortices are represented by closed
surfaces. The basic observation indicating the unity of the monopoles and
vortices is the strong correlation between the monopole trajectories and the
vortex surfaces \cite{mon_pvort,deforcrand,InterplayLat}.

Moreover, both monopole currents and P-vortices in the confinement phase
percolate. In other words, there exists a percolating (infrared, IR) cluster
which extends through the whole of the lattice. In the monopole case, it is
also well known~\cite{HT,Boyko} that apart from the percolating cluster there
exist finite (ultraviolet, UV) clusters. Observation of the finite clusters
of P-vortices was reported first in  \cite{InterplayLat}. In this note we
consider systematically the UV and IR clusters of vortices and monopoles and
discuss their interrelation. To check stability of our results against
variations in the definition of the P-vortices, we study both the direct
maximal center projection (DMCP)~\cite{greens2} and indirect maximal center
projection (IMCP)~\cite{IMC}. Preliminary results were presented in
\cite{InterplayLat}.

Traditionally, the monopoles and vortices have been considered as effective,
infrared degrees of freedom of the YM theories. One of central issues is then,
how much dominate these field fluctuations the confining potential. We will
contribute to this discussion by presenting new data on the string tension
induced by P-vortices, see Sect. 2.

More recently it was realized that the vortices possess highly non-trivial
properties in the ultraviolet as well \cite{kovalenko}. In particular, it was
found \cite{kovalenko} that the excess of the full non-Abelian action
associated with the P-vortices is ultraviolet divergent at presently
available lattices:
\beq
\label{action} S_{PV} - S_{vac}~\approx~0.54{A\over a^2}~~,
\eeq
where $A$ is the area of the vortices, $a$ is the lattice spacing and
$S_{vac}$ is the vacuum action. Thus, the probability to find a vortex of
area $A$ is suppressed exponentially  by the action factor for finite $A$ and
$a\to 0$. Nevertheless it is known \cite{mon_pvort} that the total area of
the P-vortices scales in physical units. The only interpretation of this
observation is that the suppression due to the action (\ref{action}) is
balanced by exponential enhancement due to the entropy \cite{kovalenko}.
Earlier, it was argued \cite{hidden} that the data \cite{MonAnatomy} indicate
a similar cancellation in case of the monopoles.

Thus, monopoles and vortices seem to represent a new kind of vacuum
fluctuations  which exhibit both ultraviolet and infrared scales, lattice
spacing $a$ and $\Lambda_{QCD}$, respectively. They can be called selftuned
since the action and entropy factors seem to cancel to a high degree of
accuracy without tuning any external parameter. This by itself justifies as
detailed study of the branes\footnote{By branes we mean the (infinitely) thin
vortices which carry the action (\ref{action}) and which are populated with
the monopoles.} as possible. In a way, this field theoretic facet of the
branes can be addressed even without reference to the confinement. In
particular, the branes might shed light on the dual formulation of the
Yang-Mills theories \cite{vz}.

Motivated by these considerations, we undertake here a detailed study of the
action associated with the vortices and of their geometry, the line of
investigation suggested by \cite{kovalenko}. In particular, to have an
exponentially enhanced entropy, the surface should be random on the scale of
the lattice spacing $a$. Hence, our interest is in the local geometry of the
P-vortices. A new point compared to \cite{kovalenko} is that we study the
infinite and finite clusters separately. In Sect. 3 we present results on
geometrical characteristics of the P-vortices. Sect. 4 is devoted to the
results on their action. The main definitions and details of numerical
calculations are given in the Appendix while Sect. 5 is conclusions.

\section{$Z(2)$ string tension}

\begin{figure}
\begin{center}
\includegraphics[scale=0.5,angle=270]{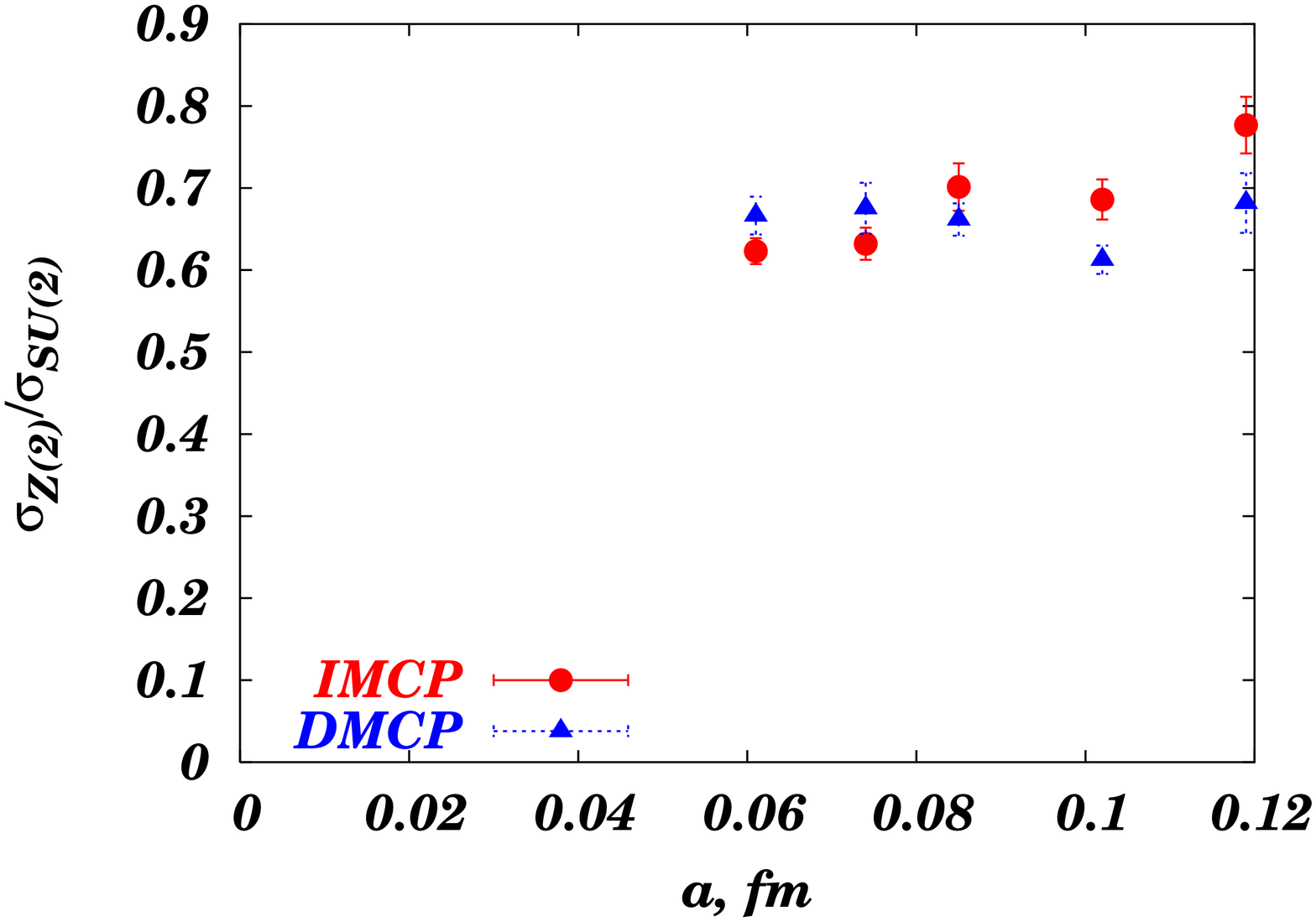}
\end{center}
\caption{The ratio of the $Z(2)$ string tension and the full $SU(2)$ string
tension for IMCP and for DMCP.
    \label{fig:str_tens}}
\end{figure}

It is well known that monopoles in the maximal Abelian projection are
responsible for about $90\%$ of the string tension both in $SU(2)$
gluodynamics~\cite{Bornyakov} and in lattice QCD with two dynamical
quarks~\cite{fullQCD}. There has been a long discussion in the literature on
whether the P-vortices reproduce  well the non-Abelian string tension (see
Refs.~\cite{IMC}, \cite{tomb2}, \cite{dramareply}, \cite{SA} and references
therein). On Fig.~\ref{fig:str_tens} we show the ratio of the $Z(2)$ string
tension obtained from Z(2) links and the full $SU(2)$ string tension. To
calculate the $Z(2)$ string tension we use the standard Creutz ratio for
loops up to $7\times 7$ lattice spacings (for all values of $\beta$). It
appears that near the continuum limit for IMCP and DMCP
$\sigma_{Z(2)}/\sigma_{SU(2)}$ is of the order of $0.6$. Probably to
reproduce the full string tension from $Z(2)$ variables we have to use some
other $Z(2)$ projection~\cite{dramareply}. In any case the fact that we
reproduce a substantial part of the string tension from $Z(2)$ variables
shows that P-vortices even in the considered projections are related to the
confinement.

\section{Geometry of P-vortices}

\subsection{Randomness of the surfaces}

\begin{figure}
\centerline{\includegraphics[scale=1., angle=0]{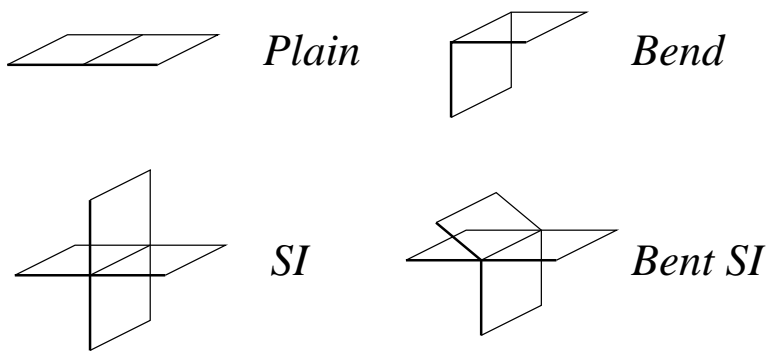}}
\caption{\label{fig:PV_junc}
    Possible two and four plaquette junctions in D=4. Junctions of
    six plaquettes are very rare, and we do not consider them.
    }
\end{figure}

\begin{figure}
\begin{center}
\includegraphics[scale=0.50, angle=270]{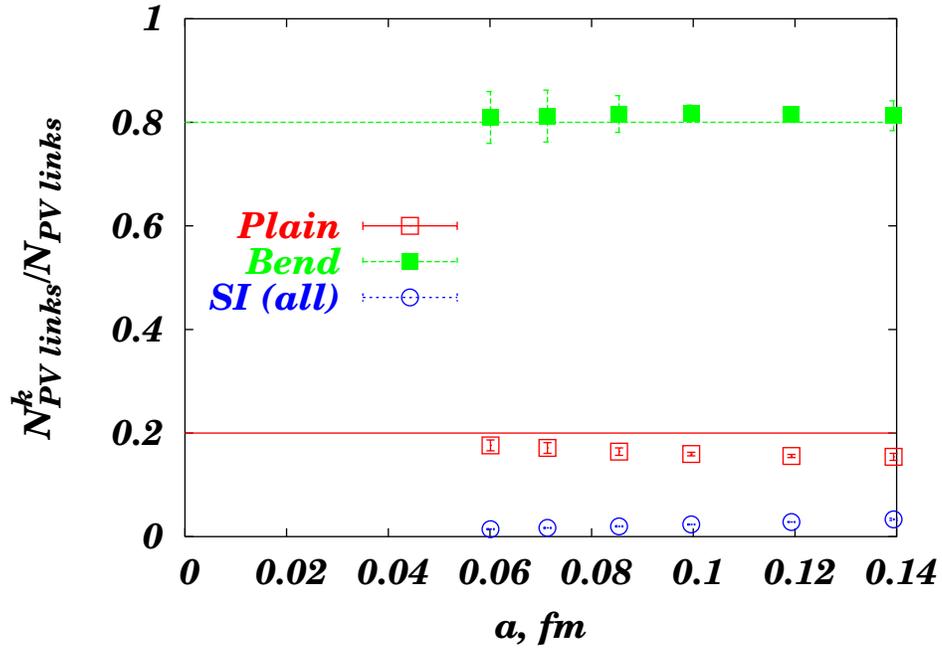}
\end{center}
\caption{\label{fig:PV_curve} Probability of different types of junctions on
the IR P-vortices for IMCP.}
\end{figure}

As is mentioned in the Introduction the selftuning of the P-vortices assumes
randomness of their surfaces at the scale of the lattice spacing $a$. Our
main point now is that the randomness can be probed through measurements. Let
us discuss first how ``random'' is the surface formed by the IR P-vortex
cluster. The plaquettes forming a surface can have the junction
configurations shown in Fig.~\ref{fig:PV_junc}. Junctions of 2 plaquettes,
can be "plain" or "bend". For a random surface in D=4 we have:
\begin{equation} \label{NpNb}
{N_{plain}\over N_{bend}} = {1\over 4}\, .
\end{equation}
If self-intersections are rare, $N_{SI}, \, N_{Bent\, SI} \ll N_{link}$
($N_{link}$ is the total number of links on the surface), then
\begin{equation}
N_{plain} = {1\over 5} N_{link}, \quad N_{bend} = {4\over 5} N_{link}\, .
\label{eqn:PV_parts}
\end{equation}
We find out  that the relations (\ref{NpNb}) and (\ref{eqn:PV_parts}) are
satisfied to a good accuracy for IR clusters in a wide region of the lattice
spacing, $a$. This is shown in Fig.~\ref{fig:PV_curve} for IMCP, for DMCP the
results are very similar. Thus, IR clusters at short distances behave like
random surfaces. Relation (\ref{eqn:PV_parts}) is also satisfied to a good
accuracy since the number of links with self intersections of P-vortices is
small ($\approx 3.4\%$ for IMCP and $\approx 2.7\%$ for DMCP for the largest
value of $a=0.139\,fm$, and even less for smaller values of the lattice
spacing).

\begin{figure}[h]
\begin{center}
\includegraphics[scale=0.50,angle=270]{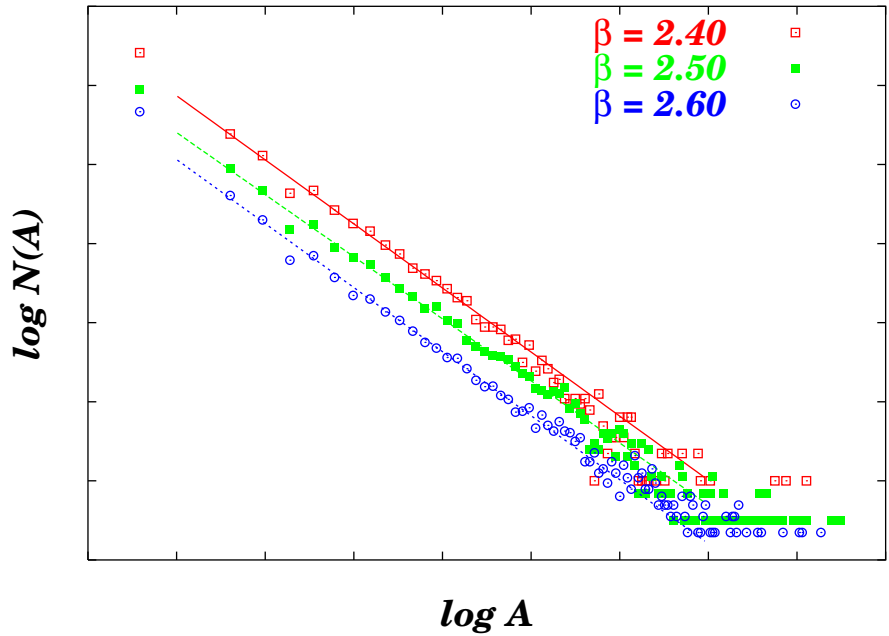}
\end{center}
\caption{UV P-vortex cluster spectrum in IMCP. Displacement in "y"-axis is made
for convenience.
    \label{fig:UVspectrum}}
\end{figure}

The ultraviolet (UV), or finite, clusters have quite different geometry. They
are dominated by small objects like $1\times 1\times 1,\;\, 1\times 1\times
2$ and so on (see Fig.~\ref{fig:UVspectrum} and sect.~\ref{sect:finite}). For
all considered lattice spacings most of plaquette junctions on UV P-vortices
clusters are "bend" (more than $95.5\%$ for IMCP and $96.4\%$ for DMCP). The
number of other plaquette junctions also weakly depends on the lattice
spacing.

\subsection{Spectrum of finite P-vortex clusters}
\label{sect:finite}

One of the basic geometric characteristics of the P-vortices is their total
area. It is known to scale in physical units~\cite{greens2,kovalenko}:
\beq\label{total}
A_{vort}~\approx ~24\,fm^{-2}\cdot V_4,
\eeq
where $V_4$ is the volume of the lattice in $fm^4$. The area (\ref{total})
includes both IR and UV clusters.

Consider now the spectrum of the UV clusters. On Fig.~\ref{fig:UVspectrum}
the number of UV clusters of the given area $A$ (lattice units) is shown in
logarithmic scale for IMCP. Three values of $\beta$ are used:
$\beta=2.40,\,\beta=2.50,\,\beta=2.60$. The fit of the spectrum by the
expression
\begin{equation}\label{tau}
N(A) \sim {1\over A^{\tau}},
\end{equation}
is shown by solid lines. The results of the fit are given in the Table~1, it is
seen that $\tau \approx 3$ for IMCP. Performing the fit we neglect the smallest
clusters (cubes with $A=6$), since they reflect the geometry of the lattice.

\begin{table}[t]\label{Atau}
\caption{Values of the parameter $\tau$}
\begin{center}
\begin{tabular}{|c|c|c|c|}
\hline $\beta$ &      $\log A_{max}$ &    $\tau$\\
 \hline
 \hline
 &  IMCP& \\
 \hline
$2.40$ &     $4.5$ & $3.24\pm 0.05$ \\
$2.50$ &     $5.5$ & $3.14\pm 0.04$ \\
$2.60$ &     $6.0$ & $3.24\pm 0.04$ \\
\hline
 &  DMCP&\\
 \hline
$2.40$ &     $5.5$ & $3.36\pm 0.06$ \\
$2.50$ &     $6.0$ & $3.49\pm 0.06$ \\
$2.60$ &     $6.0$ & $3.85\pm 0.09$ \\
\hline
\end{tabular}
\end{center}
\end{table}

To summarize, the density of the UV clusters reveals a strong dependence on
the lattice spacing $a$\footnote{The average area of UV clusters is $<A>\, =
\, \int_{a^2}^\infty N(A) A\, dA\approx \frac{C}{a^2}$}. This strong
dependence so far was consistent with the $a$ independence of the total area
(\ref{total}). Extrapolation to smaller $a$ seems, however, non-trivial.
Either (\ref{total}) or the exponential in (\ref{tau}) is to be changed at
smaller lattice spacings.

\subsection{P-vortices and monopoles}

\begin{figure}[h]
\begin{center}
\includegraphics[scale=0.5,angle=270]{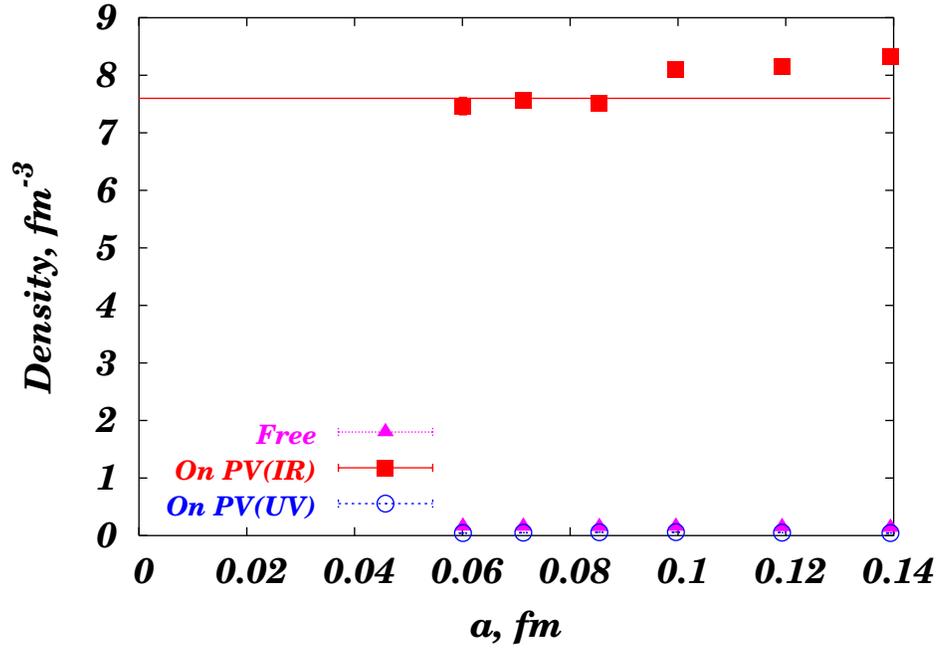}
\end{center}
\caption{Density of IR monopole currents.
    \label{fig:MonIR_on_PV_IMC}}
\end{figure}

\begin{figure}[h]
\begin{center}
\includegraphics[scale=0.5,angle=270]{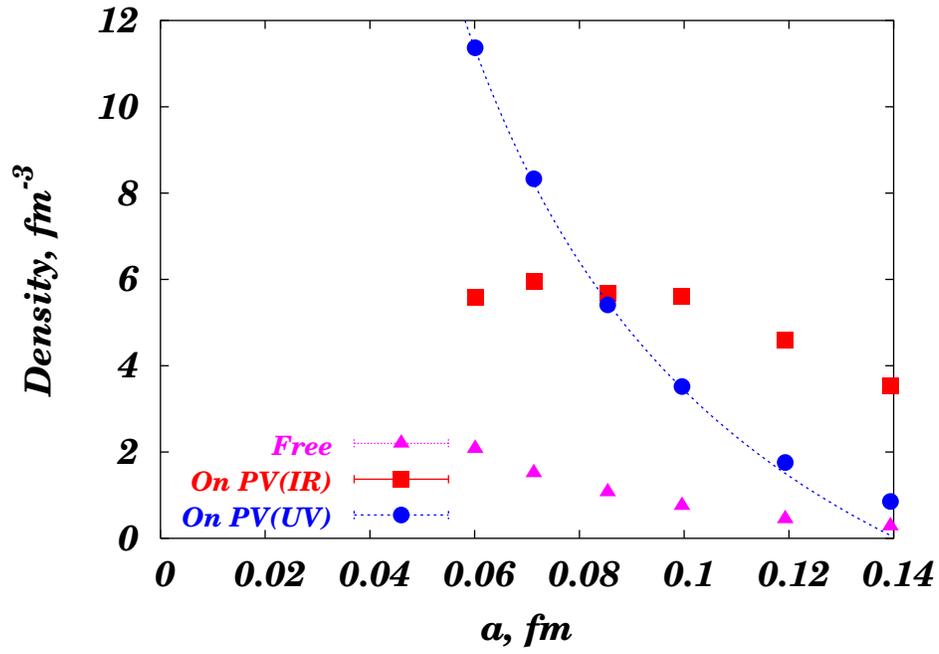}
\end{center}
\caption{Density of UV monopole currents.
    \label{fig:MonUV_on_PV_IMC}}
\end{figure}

In Refs.~\cite{mon_pvort} it was shown that at $\beta=2.4$ the main part of
the monopole trajectories extracted in the maximal Abelian projection lie on
P-vortices in IMCP. Below we show that this result is valid in a wide range
of the lattice spacing. Moreover we consider separately IR and UV clusters of
monopoles and P-vortices. In Figs.~\ref{fig:MonIR_on_PV_IMC},
\ref{fig:MonUV_on_PV_IMC} we present the densities of IR and UV monopole
clusters lying on IR and UV P-vortices in IMCP. We also show the densities of
IR and UV monopole clusters which do not belong to P-vortices (``free''
monopoles). We see that the density of IR monopoles lying on IR P-vortices is
much larger than other densities. Similarly, the density of UV monopoles is
mainly due to monopoles lying on UV P-vortices. This density is divergent as
$1/a$ at small values of $a$. The fit of four points at small values of $a$
by the function $C_1+C_2/a$ gives:
\begin{equation}
C_1\approx -8.5(2)fm^{-3},\; C_2\approx 1.19(1)fm^{-2}
\end{equation}
For DMCP we observe similar correlations of monopoles and P-vortices.

\section{Action density of P-vortices}
\subsection{Percolating cluster vs finite clusters}

As we have mentioned in the Introduction the average action associated with the
P-vortices is ultraviolet divergent, with a simple $a$ dependence, see
(\ref{action}). However, in case of the monopoles there exists also a finer
effect. Namely, the monopole action for finite (UV) clusters is somewhat higher
than for the percolating (IR) cluster \cite{MonAnatomy}. The result emphasizes
the physical nature of the lattice monopoles since it is natural that the
percolating monopoles have lower action.

\begin{figure}[h]
\begin{center}
\includegraphics[scale=0.5,angle=270]{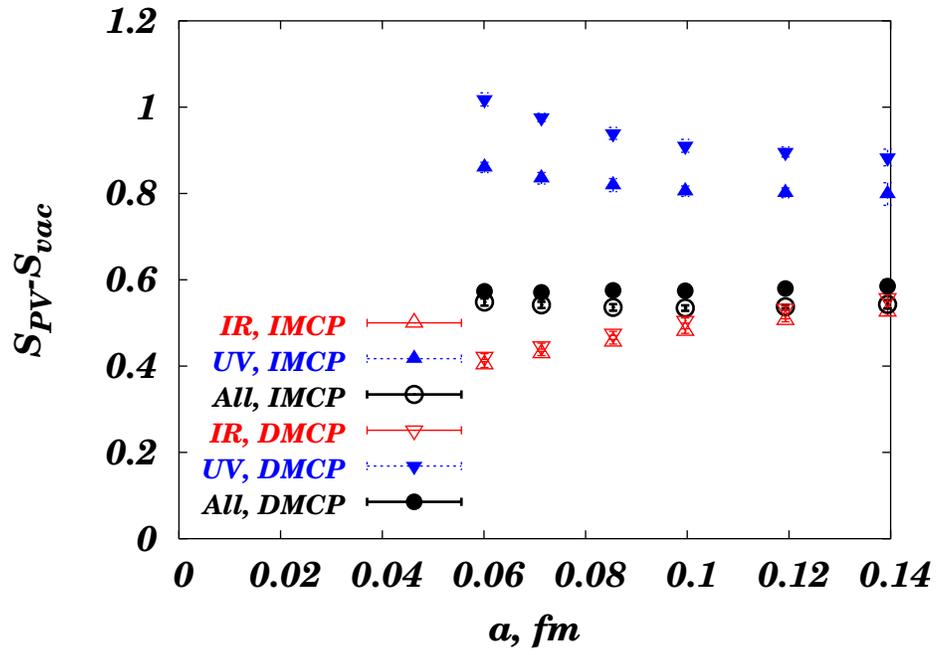}
\end{center}
\caption{\label{fig:PV_act} Excess of the plaquette action density on the
plaquettes dual to plaquettes belonging to P-vortices for IMCP and DMCP.}
\end{figure}

In this section we report on observation of a similar effect in case of the
P-vortices. Namely, in  Fig.~\ref{fig:PV_act}  we plot the excess of the
action, $S_{PV} - S_{vac}$ vs. lattice spacing $a$. The vacuum action density
is defined in the usual way, $S_{vac} = \beta <1-\frac 12 \mbox{Tr} U_P>$,
$S_{PV}$ is the same action but only for plaquettes dual to P-vortices. The
non-Abelian action density on the plaquettes dual to plaquettes belonging to
all P-vortices seems to be constant in lattice units. In other words it is
divergent in physical units as $a\to 0$. The action for IR clusters is very
close for IMCP and DMCP, while the action for UV clusters is larger for DMCP.

\subsection{Action vs local P-vortex geometry}

As mentioned in the Introduction, P-vortices appear to exhibit tuning of the
action and entropy. From theoretical point of view the very existence of such
a surface is a highly non-trivial and challenging observation. The point is
that such tuning is not possible in case of the Nambu-Goto action. Thus, one
can expect that action is actually not uniform but depends on the local
geometry.

\begin{figure}
\begin{center}
\includegraphics[scale=0.5, angle=270]{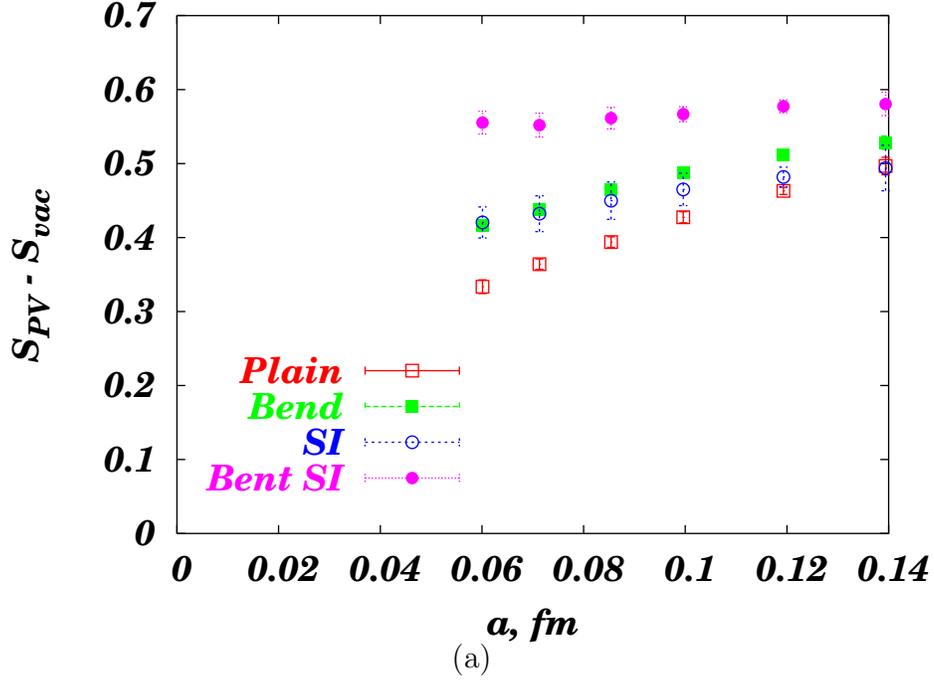}\\
(a)\\
\includegraphics[scale=0.5, angle=270]{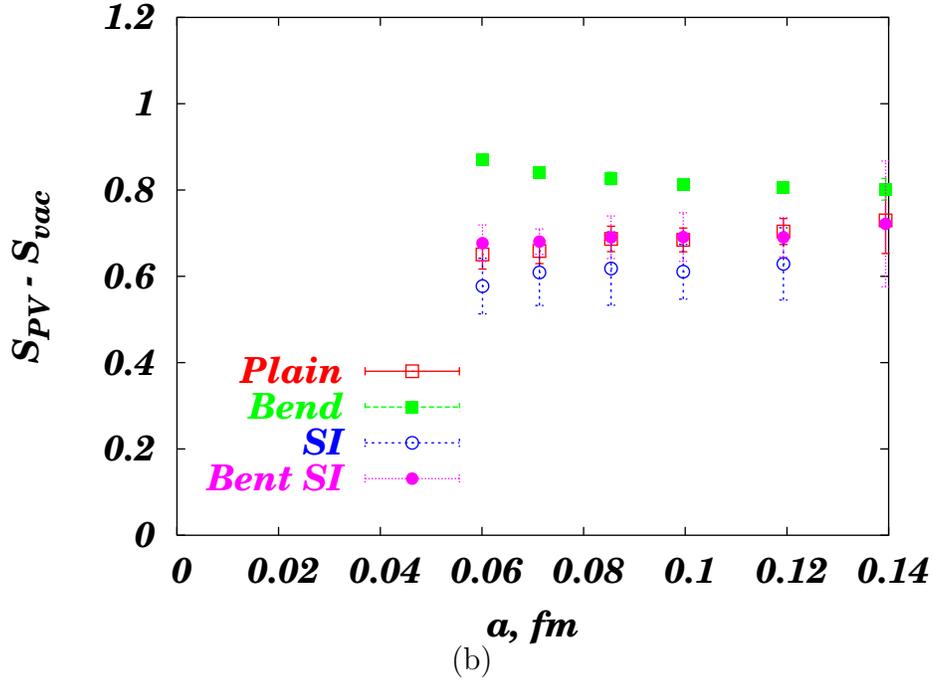}\\
(b)
\end{center}
\caption{\label{fig:PV_IMC_act_curve} Dependence of the excess of the plaquette
action on the local geometry of P-vortex cluster for IMCP for various lattice
spacings; (a) IR cluster, (b) UV cluster.}
\end{figure}

To study possible relation between the local geometry and action we have
measured dependence of the action on the type of the junction of  neighboring
plaquettes. On Fig.~\ref{fig:PV_IMC_act_curve} we show the excess of the
average plaquette action on the plaquettes which are attached to the link in
a special manner (see Fig.~\ref{fig:PV_junc}) for IMCP. For DMCP we have
analogous results. The data do indicate that the `plain' junction costs less
action than the bended ones. The tendency is especially clearly manifested in
case of the IR vortices.

\subsection{P-vortex geometry and monopoles}

\begin{figure}
\begin{center}
\includegraphics[scale=0.5, angle=270]{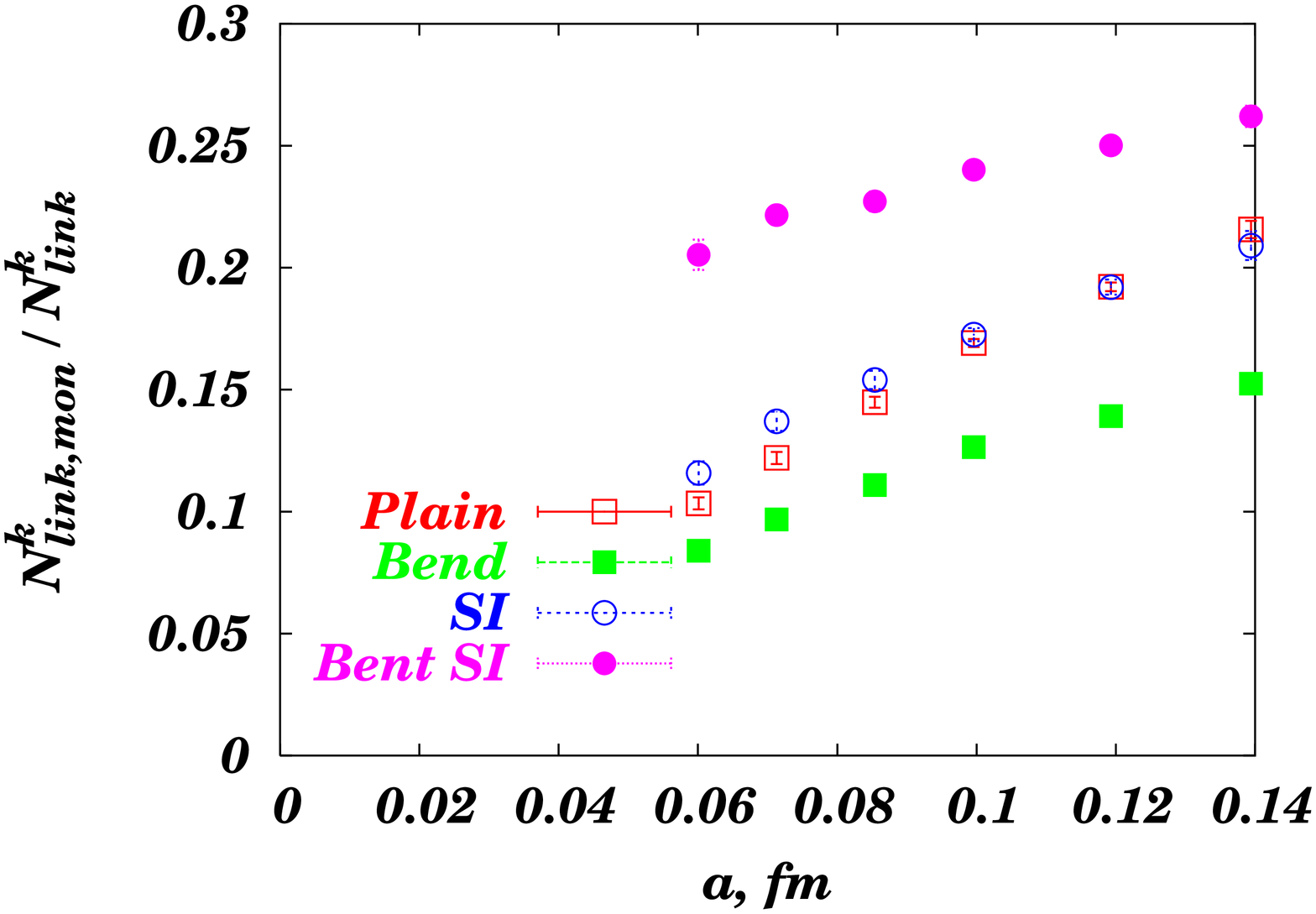}\\
(a)\\
\includegraphics[scale=0.5, angle=270]{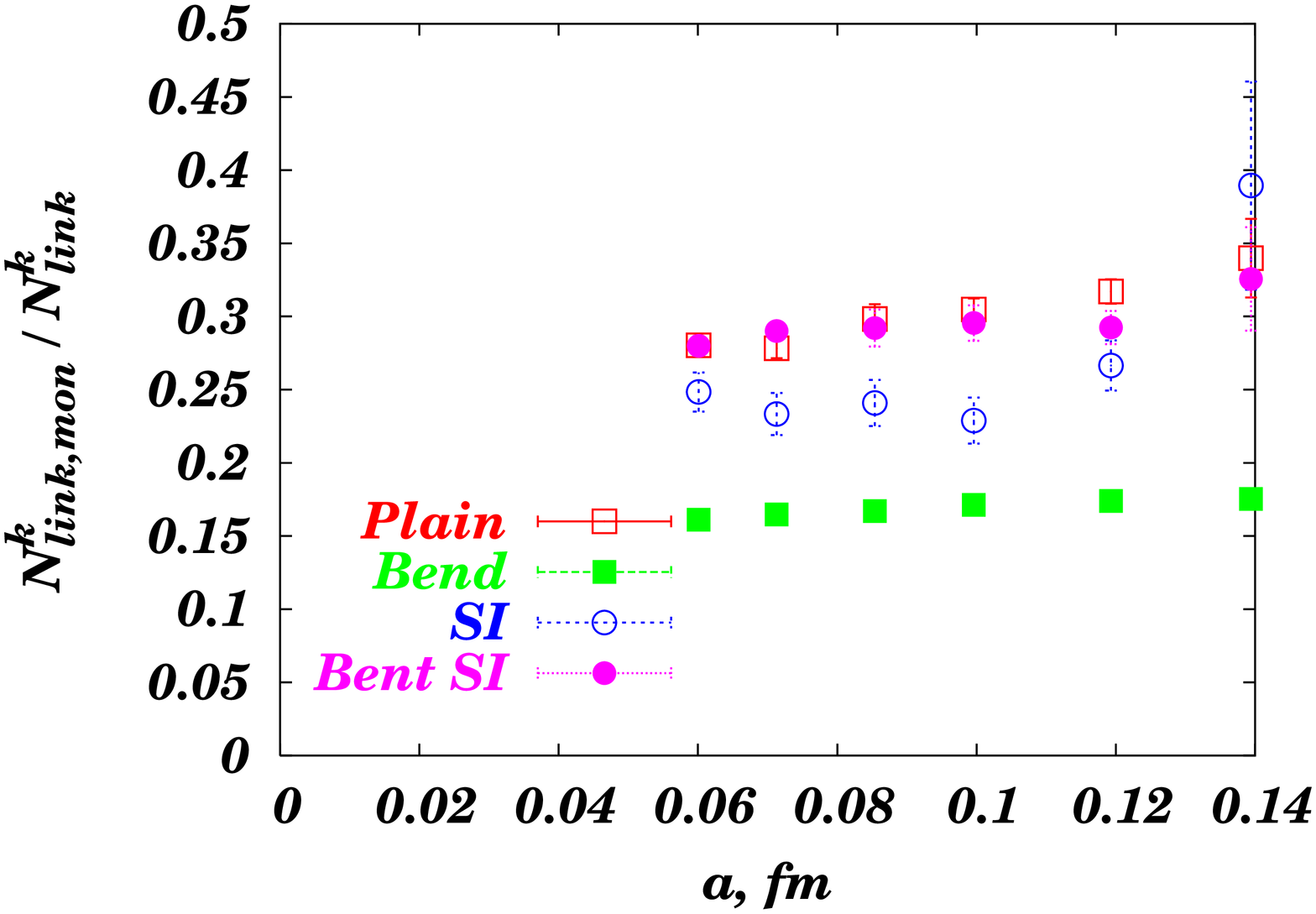}\\
(b)
\end{center}
\caption{\label{fig:PV_IMC_mon_curve} Ratio of number of links
with monopoles and number of all links of certain junction type for IMCP
P-vortices; (a) IR cluster, (b) UV cluster.}
\end{figure}

Thus, the most part of the monopole currents lie on the links which belong to
P-vortices, and now we discuss the correlation of the different types of the
plaquette junctions on P-vortex (see Fig.~\ref{fig:PV_junc}) with the
monopole currents. On Fig.~\ref{fig:PV_IMC_mon_curve} we show for IMCP the
ratio $N^{k}_{link, mon}\over N^{k}_{link}$, where $N^{k}_{link}$ is the
average number of the links corresponding to the junction $k$, $N^{k}_{link,
mon}$ is the number of links which carry monopole current and correspond to
the junction $k$. In case of DMCP, we have very similar results.

\section{Conclusions}

In this note we have presented detailed measurements of action and local
geometrical characteristics of P-vortices and their correlations with
monopoles. All the measurements are done separately for the percolating and
finite clusters. The results obtained for DMCP and IMCP are very similar.

It is worth emphasizing that all the characteristics we have been considering
are gauge invariant. Indeed, we have measured the full non-Abelian action
associated with the branes (P-vortices). Also, the geometrical characteristics
are in the physical units, or in units of $\Lambda^{-1}_{QCD}$. On the other
hand, the branes themselves are defined within a particular projection and the
definition is not unique. To reconcile these observations one is invited to
assume that through the projection one detects actually gauge invariant
objects. Then various projections are not necessarily the same effective to
detect these gauge invariant vacuum fluctuations. The criterion which worked
empirically so far is that the projections which are most effective to describe
the confining potential exhibit gauge invariant properties in the most regular
way.

Our detailed measurements did not change this picture in its basic points.
However, in quite a few cases we found substantial differences between the
properties of percolating and finite clusters of P-vortices. In
particular, it is only the surface of the percolating cluster which exhibits
the randomness on the scale of the lattice spacing $a$ (which is a prerequisite
for cancellation between the huge action and entropy factors). The small
clusters are dominated by elementary cubes. One is inclined to consider such
clusters as artifacts.

On the other hand, the total area of the infrared cluster alone does not scale
so beautifully as the total area of all the P-vortices. Also, the non-Abelian
action associated with the ultraviolet clusters of P-vortices is considerably
higher than the action for the percolating cluster. Which suggests that the
vortices are physical. Measurements at smaller $a$ are desired to distinguish
between facts and artifacts in case of the ultraviolet clusters.

One of our results is the spectrum of the ultraviolet clusters in IMCP as a
function of their area $A$,
\begin{displaymath}
N(A)~\sim~A^{-3}
\end{displaymath}
If the theory of the vortices were known, the spectrum were predictable. This
is true, in particular, in case of the monopoles, see the second paper in
Ref. \cite{hidden}. At the moment, however, the theory of the percolating
surfaces is not known yet.

We have confirmed strong correlation between the monopoles and the
P-vortices. Moreover, we have observed that the infrared monopoles are
correlated mostly with the percolating cluster of P-vortices. The
investigation of the properties of the IR monopole clusters in 4D SU(2)
lattice gauge theory allow to conclude that monopoles percolate on two
dimensional surfaces~\cite{vz}.

Detailed studies of the action of the P-vortices indicate dependence of the
action on the local geometry. Also, the monopole trajectories belonging to the
vortices are having a larger excess of the action than the vortices on average.
On the theoretical side, this observation is rather gratifying although cannot
be fully interpreted at present. Indeed, it is known (see, e.g.,
\cite{ambjorn}) that the tuning between action and entropy is not possible at
all for the simplest (Nambu-Goto) action,
\begin{displaymath}
S~=~\sigma\cdot A.
\end{displaymath}
Our results indicate that the action should include terms related to the
curvature of the surface and monopole trajectories. Both modifications of the
simplest action above have been considered on various occasions in the
literature. In particular, the action of the particles living on submanifolds
is commonly introduced in theory of D-branes, for review and references see,
e.g., \cite{dbranes}. Theories with action depending on the curvature have
been widely discussed in quantum geometry, for review and references see,
e.g. \cite{ambjorn}. Application of the latter idea to the P-vortices was
considered first in Ref. \cite{Rein}.

At present, there is no theory of P-vortices on the fundamental level.
Hopefully, the results on the action and geometry of the vortices obtained
in this paper would allow to narrow the search for
such a theory.
\section*{Acknowledgements}
A.V.K., M.I.P. and S.N.S. are partially supported by grants RFBR 02-02-17308,
RFBR 01-02-17456, DFG-RFBR 436 RUS 113/739/0, INTAS-00-00111, and CRDF award
RPI-2364-MO-02; V.I.Z. is partially supported by INTAS-00-00111 grant.

The authors are grateful to V.G.~Bornyakov and J.~Greensite for useful
discussions.

\section*{Appendix}

To define the P-vortices we use both the DMC~\cite{greens2} and the
IMC~\cite{IMC} projections. The DMCP in SU(2) lattice gauge theory is defined
by the maximization of the functional
\begin{equation}
F_1(U) = \sum_{n,\mu} \left( \tr U_{n,\mu}\right)^2 \, , \label{maxfunc}
\end{equation}
with respect to gauge transformations, $U_{n,\mu}$ is the lattice gauge
field. The maximization of (\ref{maxfunc}) fixes the gauge up to Z(2) gauge
transformations and the corresponding Z(2) gauge field is defined as:
$Z_{n,\mu} = \mbox{sign} \tr U_{n,\mu}$. The plaquettes $Z_{n,\mu\nu}$
constructed as product of links $Z_{n,\mu}$ along the border of the plaquette
have values $\pm 1$. The P-vortices (forming closed surfaces in 4D space) are
made from the plaquettes, dual to plaquettes with $Z_{n,\mu\nu} = -1$.

To get IMCP we first fix the maximally Abelian gauge by maximizing the
functional
\begin{equation}
F_2(U) = \sum_{n,\mu} \tr \left( U_{n,\mu}\sigma_3 U_{n,\mu}^+ \sigma_3\right)
\, , \label{maxfuncmaa}
\end{equation}
with respect to gauge transformations. This procedure leaves unfixed $U(1)$
degrees of freedom, the corresponding $U(1)$ compact gauge field is
$e^{i\theta_{n,\mu}}$, $\theta_{n,\mu}$ being the phase of the $(1,1)$
element of the link matrix $U_{n,\mu}$. After that we can extract monopole
currents from the Abelian fields. Finally, we project gauge degrees of
freedom $U(1)\to Z(2)$ by the procedure analogous to the DMCP case. We
substitute into eq.~\eq{maxfunc} the Abelian matrix: $U_{n,\mu} \to
U_{n,\mu}^{Ab}\equiv \mbox{diag}\left( {e^{i\theta_{n,\mu}},
e^{-i\theta_{n,\mu}}}\right)$ and maximize $F_1(U)$ with respect to $U(1)$
gauge transformations.

We work at various lattice spacings to check the existence of the continuum
limit of our observables. The parameters of our gauge field configurations are
listed in Table~2. To fix the physical scale we use the string tension in
lattice units~\cite{Fingberg:1992ju}, $\sqrt\sigma = 440\, MeV$.

\begin{table}[t]\label{conf_table}
\caption{Parameters of configurations.}
\begin{center}
\begin{tabular}{|c|c|c|c|}
\hline
$\beta$ & Size &    $N_{IMCP}$ &    $N_{DMCP}$ \\
\hline
$2.35$ &    $16^4$ &    $20$ &  $20$ \\
$2.40$ &    $24^4$ &    $50$ &  $20$ \\
$2.45$ &    $24^4$ &    $20$ &  $20$ \\
$2.50$ &    $24^4$ &    $50$ &  $20$ \\
$2.55$ &    $28^4$ &    $37$ &  $17$ \\
$2.60$ &    $28^4$ &    $50$ &  $20$ \\
\hline
\end{tabular}
\end{center}
\end{table}

To fix the maximally Abelian gauge and maximal center gauge we create 20
randomly gauge transformed copies of the gauge field configuration and apply
the Simulated Annealing~\cite{Bornyakov,SA} algorithm to each copy. We use in
calculations that copy which correspond to the maximal value of the gauge
fixing functional. To fix the indirect maximal center gauge from
configuration fixed to maximally Abelian gauge one gauge copy is enough to
work with our accuracy.

\newpage


\begin{thebibliography}{99}

\bibitem{reviews}
M.N.~Chernodub, M.I.~Polikarpov, in {\it 'Cambridge 1997, Confinement, duality,
and nonperturbative aspects of QCD'},
p.~387; {\tt hep-th/9710205};\\
 J.~Greensite, {\it  Prog. Part. Nucl. Phys.}
{\bf 51}, 1 (2003), {\tt hep-lat/0301023}

\bibitem{mon_pvort}
L.~Del~Debbio, M.~Faber, J.~Greensite, S.~Olejnik, in 'Zakopane 1997, New
developments in quantum field theory', p. 47, {\tt hep-lat/9708023};\\
J.~Ambjorn, J.~Giedt, J.~Greensite, {\it JHEP} {\bf 0002}, 033 (2000).

\bibitem{deforcrand}  P.~de~Forcrand, M.~Pepe, {\it Nucl.\ Phys.}\
{\bf B598}, 557 (2001).

\bibitem{InterplayLat}
A.~V.~Kovalenko, M.~I.~Polikarpov, S.~N.~Syritsyn and V.~I.~Zakharov, {\it
``Interplay of monopoles and P-vortices''}, preprint ITEP-LAT/2003-23, {\tt
hep-lat/0309032}.

\bibitem{HT}
A.~Hart and M.~Teper, {\it Phys. Rev.} D58, 014504 (1998).

\bibitem{Boyko}  V.G.~Bornyakov, P.Yu.~Boyko, M.I.~Polikarpov and
V.I.~Zakharov, {\it Nucl. Phys.} {\bf B672} 222 (2003).

\bibitem{greens2} L.~Del~Debbio, M.~Faber, J.~Giedt, J.~Greensite and
S.~Olejnik, {\it Phys.Rev.} {\bf D58}, 094501 (1998).

\bibitem{IMC}
L.~Del~Debbio, M.~Faber, J.~Greensite, S.~Olejnik, {\it Phys. Rev.} {\bf D55},
2298 (1997).

\bibitem{kovalenko}
F.~V.~Gubarev, A.~V.~Kovalenko, M.~I.~Polikarpov, S.~N.~Syritsyn and
V.~I.~Zakharov, {\it Phys. Lett.} {\bf B574}, 136 (2003).

\bibitem{hidden}
V.~I.~Zakharov, {\it ``Hidden mass hierarchy in QCD''}, {\tt hep-ph/0204040};\\
M.~N.~Chernodub and V.~I.~Zakharov, {\it Nucl. Phys.} {\bf B669}, 233 (2003).


\bibitem{MonAnatomy}
V.~G.~Bornyakov, M.~N.~Chernodub, F.~V.~Gubarev, M.~I.~Polikarpov,
T.~Suzuki, A.~I.~Veselov and V.~I.~Zakharov,
{\it Phys. Lett.} {\bf B537}, 291 (2002).


\bibitem{vz}
V.~I.~Zakharov,
{\it ``Hints on dual variables from the lattice $SU(2)$ gluodynamics''},
{\tt hep-ph/0309301}.


\bibitem{Bornyakov}
G.~S.~Bali, V.~Bornyakov, M.~Muller-Preussker and K.~Schilling, {\it Phys.
Rev.} {\bf D54}, 2863 (1996).

\bibitem{fullQCD}
DIK Collaboration (V.~G.~Bornyakov et al.), {\it ``Dynamics of monopoles and
flux tubes in two flavor dynamical QCD''}, {\tt hep-lat/0310011}.

\bibitem{tomb2}
T.~G.~Kovacs and E.~T.~Tomboulis, {\it Phys. Lett.} {\bf B463}, 104 (1999).

\bibitem{dramareply}
M.~Faber, J.~Greensite and S.~Olejnik, {\it Phys. Rev.} {\bf D64}, 034511
(2001).

\bibitem{SA}
V.~G.~Bornyakov, D.~A.~Komarov and M.~I.~Polikarpov,
{\it Phys. Lett.} {\bf B497} 151 (2001);\\
V.~G.~Bornyakov, D.~A.~Komarov, M.~I.~Polikarpov and A.~I.~Veselov,
in 'Osaka 2000,
{\it ``Quantum chromodynamics and color confinement''}, p.133',
{\tt hep-lat/0210047}.


\bibitem{ambjorn}
J.~Ambjorn, {\it ``Quantization of geometry''}, {\tt hep-th/9411179}.

\bibitem{dbranes}
C.~V.~Johnson, ``D-Branes'', Cambridge monographs on mathematical physics,
Cambridge (2003).

\bibitem{Rein}
M.~Engelhardt, H.~Reinhardt, {\it Nucl.Phys.} {\bf B585}, 591 (2000).

\bibitem{Fingberg:1992ju}
J.~Fingberg, U.~M.~Heller and F.~Karsch, {\it Nucl. Phys.} {\bf B392}, 493
(1993).

\end{thebibliography}
\end{document}